Article

# The NuGrid AGB Evolution and Nucleosynthesis Data Set

Umberto Battino [1,*,†], Marco Pignatari [1,2,3,†], Ashley Tattersall [4,†], Pavel Denissenkov [3,5,†] and Falk Herwig [3,5,†]

1. E.A. Milne Centre for Astrophysics, Department of Physics and Mathematics, University of Hull, Hull HU6 7RX, UK; mpignatari@gmail.com
2. Konkoly Observatory, Research Centre for Astronomy and Earth Sciences, Eötvös Loránd Research Network (ELKH), Konkoly Thege M. út 15-17, 1121 Budapest, Hungary
3. Joint Institute for Nuclear Astrophysics, Center for the Evolution of the Elements, Michigan State University, 640 South Shaw Lane, East Lansing, MI 48824, USA; pavelden@uvic.ca (P.D.); fherwig@uvic.ca (F.H.)
4. School of Physics and Astronomy, University of Edinburgh, Edinburgh EH9 3FD, UK; ashley.tattersall@ed.ac.uk
5. Department of Physics & Astronomy, University of Victoria, Victoria, BC V8P 5C2, Canada
* Correspondence: u.battino@hull.ac.uk
† The NuGrid Collaboration, http://www.nugridstars.org (accessed on 1 March 2022).

**Abstract:** Asymptotic Giant Branch (AGB) stars play a key role in the chemical evolution of galaxies. These stars are the fundamental stellar site for the production of light elements such as C, N and F, and half of the elements heavier than Fe via the slow neutron capture process (*s*-process). Hence, detailed computational models of AGB stars' evolution and nucleosynthesis are essential for galactic chemical evolution. In this work, we discuss the progress in updating the NuGrid data set of AGB stellar models and abundance yields. All stellar models have been computed using the MESA stellar evolution code, coupled with the post-processing mppnp code to calculate the full nucleosynthesis. The final data set will include the initial masses $M_{ini}/M_\odot$ = 1, 1.65, 2, 3, 4, 5, 6 and 7 for initial metallicities Z = 0.0001, 0.001, 0.006, 0.01, 0.02 and 0.03. Observed *s*-process abundances on the surfaces of evolved stars as well as the typical light elements in the composition of H-deficient post-AGB stars are reproduced. A key short-term goal is to complete and expand the AGB stars data set for the full metallicity range. Chemical yield tables are provided for the available models.

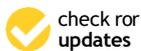

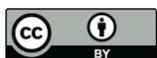

## 1. Introduction

The abundance distribution in the solar system is the result of the nucleosynthesis in several generations of stars [1–5]. Metallicity-dependent stellar yields from theoretical stellar models are a fundamental ingredient of galactic chemical evolution, and are used to study how the abundances have formed in the Sun and in other stars in the Galaxy. Stars with different initial masses and metallicities contribute in different ways to the production of elements. Compared to massive stars, low- and intermediate-mass stars (M ≲ 8M$_\odot$) contribute to the chemical evolution of the interstellar medium over longer timescales. These types of stars finish their evolution as a compact white dwarf, and predominantly eject nucleosynthesis products via stellar winds during the asymptotic giant branch (AGB) phase [6–8]. In case they are part of a binary system, depending on the system configuration more elements can be made by these stars via nova events [9–11] and other variable objects, such as R Coronae Borealis stars [12,13], or by Type Ia Supernovae [14–17]. Depending on the initial stellar mass, during the AGB phase, light elements such as carbon, nitrogen, fluorine and sodium can be made by charged particle reactions activated at the bottom of the AGB envelope or in the He intershell region immediately below [18–20]. Additionally, AGB stars are also a fundamental source of heavy elements beyond iron via the slow neutron-capture process (*s*-process [21–23]). The *s*-process occurs in a succession of neutron captures whose typical timescales are longer than the beta-decay half-life of the unstable isotopes produced. As a result, the reaction path proceeds along the valley of stability. Most of the s elements between Fe and Sr (60 < A < 90) in the Sun are produced by the weak *s*-process component in massive stars [23–25]. The effective contribution in the same mass region from massive AGB stars and super-AGB stars is quite uncertain. For A > 90, *s*-process elements are mostly produced by low-mass AGB stars [3,26,27].

After low- and intermediate-mass stars have consumed all H and He in their cores, during the AGB phase they burn alternatively H and He in thin shells around an inert CO core [7]. As a consequence of recurrent thermal pulses driven by He fusion, during the third dredge-up phase, protons diffuse from the convective H envelope into the He intershell, allowing proton captures on the abundant $^{12}$C. In low-mass AGB stars, this allows the formation of a radiative $^{13}$C-pocket [28]. Here most of the *s*-process material is made via the $^{13}$C($\alpha$,n)$^{16}$O reaction [21,28–34]. An additional neutron source is given by the $^{22}$Ne($\alpha$,n)$^{25}$Mg nuclear reaction, which is activated at much higher temperature (T > $2 \times 10^8$ K) during the He-shell flash at the bottom of the thermal-pulse-driven convection zone [21]. The $^{22}$Ne($\alpha$,n)$^{25}$Mg is instead the dominant neutron source for the *s*-process in massive AGB and super-AGB stars [8]. Note however that the activation of the intermediate neutron capture process (*i* process [35]) could significantly affect the production of heavy elements in super AGB stars, with the activation of the $^{13}$C($\alpha$,n)$^{16}$O [36].

The products of the *s*-process have been directly observed using spectroscopy measurements for AGB stars [22,37,38], planetary nebulae [39,40], post-AGB stars [41], as well as in presolar grains condensed in the winds of old AGB stars and extracted from meteorites in the Solar System [42–46]. The comparison of stellar AGB models with these direct observations provides independent constraints for theoretical simulations and for their stellar yields used in GCE models [4,26]. The NuGrid collaboration provides grids of stellar models and yields to use for these studies, including AGB stellar models [47–49]. Ritter et al. [48] presented a set of stellar models and nucleosynthesis data, covering metallicities between Z = 0.0001 and Z = 0.02 and initial masses between 1$M_\odot$ and 25$M_\odot$. However, for low-mass AGB stars, a low *s*-process production was obtained compared to the most *s*-process-rich stars and the bulk of *s*-process isotopic measurements from presolar grains [47]. This was due to the small $^{13}$C-pocket (2–3 $10^{-5}M_\odot$) obtained in those models, where convective boundary mixing (CBM) at the bottom of the convective envelope during the third dredge up was calibrated for low-mass stellar models based on [42]. Battino et al. [33] calculated new AGB models where CBM at the bottom envelope during the third dredge-up is instead induced by internal gravity waves (IGWs), as described by [50]. This results in a 3–4 times larger radiative $^{13}$C-pocket, increasing the total *s*-process production by about the same factor. The same recipe was adopted by [49], who produced a set of low-mass AGB stellar models around solar metallicity and extending the dataset by [48] to include the supersolar initial metallicity Z = 0.03. An additional extension of the set of the new AGB models was presented in [51], focused on low-metallicity (Z = 0.001, 0.002), low-mass AGB models. In particular, from these models, it is shown that stellar models with a $^{13}$C-pocket size of at least $\sim 3 \times 10^{-4} M_\odot$ (see also, e.g., [21,42,52]), and with very low mixing below the He intershell region during helium flashes [53], are favored when compared to more *s*-process-rich observations.

In this work, we discuss the initial steps in building the new combined NuGrid data set of AGB stellar models and abundance yields. In Section 2 we describe the computational tools and nuclear networks adopted. The recommended NuGrid stellar models and ejected yields data set are discussed in Sections 3 and 4, respectively. Our conclusions and future plans are presented in Section 5.

## 2. Computational Methods

The structure of all models described in this work was computed using the stellar code MESA (revision 3709 [54]). For the initial composition, a solar-scaled and an alpha-enhanced solar distribution for models at Z $\geq$ 0.01 and Z $\leq$ 0.006 were used, respectively, based on [55], which implies a solar metallicity Z = 0.018. The modeling assumptions are summarized in [48,49,51].

The complete nucleosynthesis to generate the abundance yields during the evolution of each model is made by the multizone postprocessing frame mppnp of the NuGrid postprocessing code [47,48]. The nuclear network setup used for the calculations increases dynamically as needed, up to a limit of 5234 isotopes and with 74,313 reactions. In the case of AGB nucleosynthesis, the network never extends far from the valley of stability and neutron-rich stable isotopes are mostly made, limiting the typical network size well within the allowed capability. The NuGrid physics package uses nuclear data from multiple sources, including both nuclear physics compilations and individual experimental rates. The nuclear rates used for these simulations are summarized in [48,49]. Considering the nuclear reactions most relevant for the *s*-process, the $^{13}$C($\alpha$,n)$^{16}$O nuclear reaction rate is taken from [56]. Low-mass AGB models with initial mass M = 2M$_\odot$ and 3M$_\odot$ at around solar metallicity (Z = 0.01, 0.02, 0.03; presented in [49]) were computed using the $^{22}$Ne($\alpha$,n)$^{25}$Mg and $^{22}$Ne($\alpha$,$\gamma$)$^{26}$Mg nuclear reaction rates from [57] and NACRE [58], respectively, while for low metallicity models (Z = 0.001, 0.002; presented in [51]) with the same initial masses the more recent reaction rates from [59] were adopted. In order to keep the internal consistency across the whole metallicity range, we started recalculating our nucleosynthesis results for low-mass AGB models at around solar metallicity using the $^{22}$Ne($\alpha$,n)$^{25}$Mg and $^{22}$Ne($\alpha$,$\gamma$)$^{26}$Mg nuclear reaction rates from [59], and the same procedure will be extended to all other stellar models in the future. Neutron capture reaction rates are taken from the KADoNIS compilation [60]. For rates not included in KADoNIS, we adopt theoretical data from the JINA-REACLIB database revision 1.1 [61]. The $\beta$-decay rates are from [62] or [63] for light species and from [64,65] for the iron group and for species heavier than iron.

In massive AGB stars and super-AGB stars, hot-bottom burning (HBB) is occurring at the bottom of the AGB envelope [20,66–68]. In our postprocessing models, HBB nucleosynthesis needs to be resolved using computation timesteps smaller than the convective turnover timescale of the AGB envelope (in the order of an hour). As described in [48], for these special conditions we adopted a nested-network postprocessing approach, in which mixing and burning operators are coupled. The nested network includes the Cameron–Fowler transport mechanism to simulate the production of $^7$Li [69], the CNO, NeNa, and MgAl cycles and isotopes up to $^{35}$Cl [48]. In short, this small network is solved for zones of the convective envelope and the large decoupled network for the whole stellar model. After each time step, the abundances from the coupled solution replace the abundances from the large network.

## 3. Recommended Stellar Models Data Set

In Table 1, the present status of the new NuGrid AGB dataset is outlined. Low-mass AGB models at with low metallicity are from [49,51], respectively. The models with initial mass M = 1 and 1.65M$_\odot$ and for all metallicities are from [48]. The available massive AGB models are from [48].

The nucleosynthesis yields of [49] AGB models are recalculated in this work, adopting the $^{22}$Ne($\alpha$,n)$^{25}$Mg and $^{22}$Ne($\alpha$,$\gamma$)$^{26}$Mg reaction rates from [59]. We compare our results with key observables in the next section. The m2z2m3-bigpoc model (M$_{ini}$/M$_\odot$ = 2; Z = 0.002) from [51] is not included in the table, as a significant update of it will be presented in a dedicated forthcoming paper.

**Table 1.** The recommended AGB NuGrid dataset.

| $Z_{ini}$ | $M_{ini}/M_\odot$ | Reference | Notes |
|---|---|---|---|
| 0.03 | 2, 3 | [49] | Repostprocessed with $^{22}$Ne+$\alpha$ from [59] |
| 0.02 | 1, 1.65 | [48] | - |
| 0.02 | 2, 3 | [49] | Repostprocessed with $^{22}$Ne+$\alpha$ from [59] |
| 0.02 | 4, 5, 6, 7 | [48] | - |
| 0.01 | 1, 1.65 | [48] | - |
| 0.01 | 2, 3 | [49] | Repostprocessed with $^{22}$Ne+$\alpha$ from [59] |
| 0.01 | 4, 5, 6, 7 | [48] | - |
| 0.001 | 1, 1.65 | [48] | - |
| 0.001 | 2, 3 | [51] | $^{22}$Ne+$\alpha$ from [59] |
| 0.001 | 4, 5, 6, 7 | [48] | - |
| 0.0001 | 1, 1.65, 4, 5, 6, 7 | [48] | - |

### 3.1. Low-Mass AGB stars

The nucleosynthesis of the AGB model m3z3m2-hCBM ($M_{ini}/M_\odot$ = 3; Z = 0.03) from [49] was recomputed by [59], who showed that their newly determined $^{22}$Ne($\alpha$,n)$^{25}$Mg and $^{22}$Ne($\alpha,\gamma$)$^{26}$Mg rates remarkably improve the comparison with both surface abundances on C-rich stars and presolar grains. In Figure 1, we compare our new nucleosynthesis results of stellar models m3z3m2-hCBM and m3z2m2-hCBM ($M_{ini}/M_\odot$ = 3; Z = 0.02) from [49] with Ba isotopic ratios from presolar SiC grains from [45,46]. It is visible how the lower reaction rates from [59], compared to the ones from [57], bring the theoretical predictions to a much better agreement with observations, a larger fraction of the observed $^{138}$Ba/$^{136}$Ba values range to be found between the two red tracks. This means that grains data are consistent with predictions of models, with initial metallicities ranging between Z = 0.02 and Z = 0.03. In particular, the lower contribution from the $^{22}$Ne($\alpha$,n)$^{25}$Mg reaction causes a reduced production of $^{138}$Ba, remarkably improving the comparison with grains data.

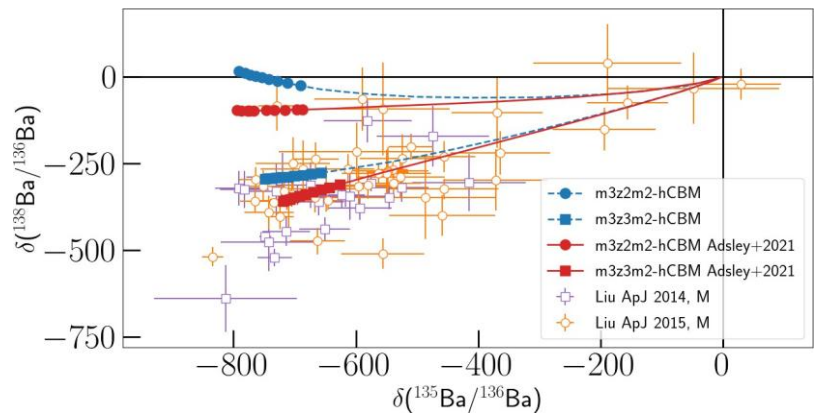

**Figure 1.** Comparison of measured Ba isotopic ratios from presolar SiC grains with the results of [49] (in blue) and the same stellar models using the $^{22}$Ne+$\alpha$ reaction rates from [59] (in red). Each circle/square marks a third dredge-up event during the C-rich AGB phase, which progressively enriches the stellar envelope in *s*-process elements.

In Figure 2 we compare the trans-Fe nucleosynthesis results from the Z = 0.001 low-mass AGB models in Table 1 with surface abundances of Ba stars HD 123396 [70] ([Fe/H] = −1.04) and HD 123396 [71] ([Fe/H] = −0.82). In particular, our 3$M_\odot$ model would still be consistent with all observed abundances within uncertainties if a dilution factor of about 0.2 dex is applied. Considering as an example a 2$M_\odot$ star companion, this translates into roughly 0.8$M_\odot$ accreted from the primary star. In general, the internal mixing parameters setting chosen to form the $^{13}$C-pocket for these AGB models allows a good agreement to be obtained between theoretical predictions and observed abundances.

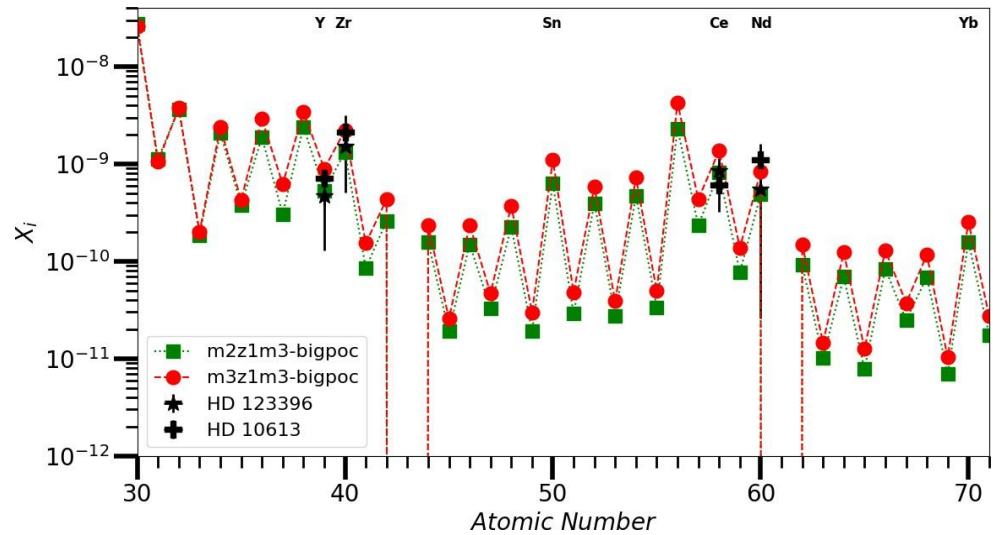

**Figure 2.** Heavy-element mass fractions of stellar models m2z1m3-bigpoc (2M$_\odot$, Z = 0.001, green squares) and m3z1m3-bigpoc (3M$_\odot$, Z = 0.001, red circles) from [51]. Surface abundances and related uncertainties of Ba stars HD 123396 [70] and HD 10613 [71] for two first-peak (Y and Zr) and second-peak elements (Ce and Nd) are also presented.

*3.2. Massive AGB Stars*

All the recommended massive AGB models in Table 1 are from [48]. In Figure 3 we show the element overproduction factors for the M = 5M$_\odot$ models from [48], which is representative of all our massive AGB models at all the metallicities listed in Table 1. Li is efficiently produced during HBB via the Cameron–Fowler mechanism at the hot bottom of the convective envelope and the decay of $^7$Be into $^7$Li in cooler outer layers [66,69]. Additionally, HBB converts $^{12}$C into $^{14}$N through a partial CN cycle, resulting in the synthesis of large amounts of N in the final ejected yields. The production of N increases in stellar models at lower metallicity due to the higher temperatures (and hence a more efficient CN cycle) at the bottom of the convective envelope. We finally notice how also Rb is overproduced compared to nearby elements. However, as discussed in [48], with present models the contribution of massive AGB stars to the production of Rb is smaller compared to low-mass AGB stars at all metallicities.

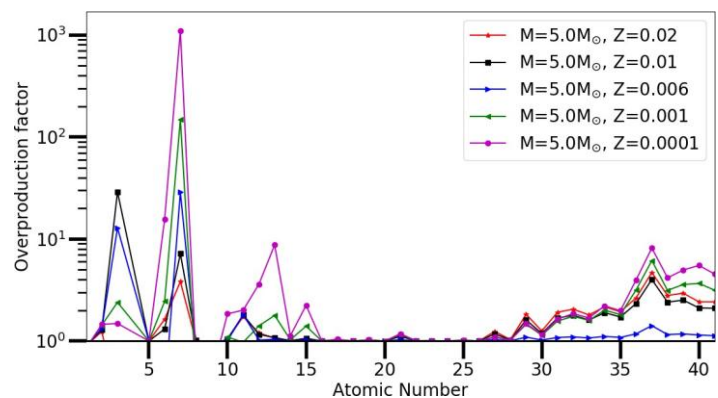

**Figure 3.** Element overproduction factors in the ejected yields for the M = 5M$_\odot$ models from [48].

## 4. Ejected Stellar Yields

We have calculated the nucleosynthesis-ejected yields for all the NuGrid models. These are shown in Table 2, where the recommended yields set from the 3M$_\odot$ models for selected isotopes are presented. The largest amount of $^{13}$C is ejected by the low-metallicity stellar model m3z1m3-bigpoc. This is a consequence of the more extended $^{13}$C-pocket formed in low-metallicity simulations compared to around-solar-metallicity ones, as discussed in [51]. As expected for trans-Fe elements, as the neutron exposure increases with decreasing metallicity, the ejected abundances distribution changes from being peaked at first peak elements for Z = 0.03 and 0.002 models, to a flat one across all three peaks at Z = 0.01, and finally to being peaked at lead at Z = 0.001. Full tables of ejected yields, from all models listed in Table 1 are available online in the NuGrid data repository at https://nugrid.github.io/content/data (accessed on 1 March 2022). NUGrid data can be explored interactively on the Astrohub platform at https://astrohub.uvic.ca/ (accessed on 1 March 2022) via the Web Exploration of NuGrid Data Interactive (WENDI). Listing 1 is an example code to analyse NuGrid's 2M$_\odot$, Z = 0.02 model and make an abundance profile plot in a WENDI Astrohub python notebook.

**Table 2.** Comparison for selected isotopes between the yields in solar masses ejected by the 3M$_\odot$ models at different metallicities in Table 1. Full ejected yields tables for all models are available online at https://nugrid.github.io/content/data (accessed on 1 March 2022).

| Isotope | Z = 0.03 | Z = 0.02 | Z = 0.01 | Z = 0.001 |
|---|---|---|---|---|
| C 12  | $4.126 \times 10^{-2}$ | $3.728 \times 10^{-2}$ | $3.083 \times 10^{-2}$ | $4.481 \times 10^{-3}$ |
| C 13  | $3.071 \times 10^{-4}$ | $2.227 \times 10^{-4}$ | $1.104 \times 10^{-4}$ | $4.974 \times 10^{-4}$ |
| N 14  | $9.125 \times 10^{-3}$ | $7.024 \times 10^{-3}$ | $3.740 \times 10^{-3}$ | $7.028 \times 10^{-4}$ |
| O 16  | $4.246 \times 10^{-2}$ | $3.414 \times 10^{-2}$ | $2.354 \times 10^{-2}$ | $1.969 \times 10^{-3}$ |
| F 19  | $6.240 \times 10^{-6}$ | $4.595 \times 10^{-6}$ | $2.762 \times 10^{-6}$ | $8.850 \times 10^{-7}$ |
| Ne 20 | $5.733 \times 10^{-3}$ | $4.221 \times 10^{-3}$ | $2.128 \times 10^{-3}$ | $1.229 \times 10^{-4}$ |
| Ne 22 | $5.027 \times 10^{-3}$ | $3.480 \times 10^{-3}$ | $1.929 \times 10^{-3}$ | $1.571 \times 10^{-4}$ |
| Na 23 | $2.517 \times 10^{-4}$ | $1.718 \times 10^{-4}$ | $8.325 \times 10^{-5}$ | $2.319 \times 10^{-6}$ |
| Mg 24 | $1.844 \times 10^{-3}$ | $1.352 \times 10^{-3}$ | $6.738 \times 10^{-4}$ | $3.222 \times 10^{-5}$ |
| Mg 25 | $3.039 \times 10^{-4}$ | $2.407 \times 10^{-4}$ | $1.252 \times 10^{-4}$ | $2.289 \times 10^{-6}$ |
| Mg 26 | $3.699 \times 10^{-4}$ | $2.838 \times 10^{-4}$ | $1.553 \times 10^{-4}$ | $3.554 \times 10^{-6}$ |
| Al 26 | $4.262 \times 10^{-7}$ | $1.182 \times 10^{-7}$ | $7.119 \times 10^{-8}$ | $2.598 \times 10^{-9}$ |
| Al 27 | $2.076 \times 10^{-4}$ | $1.523 \times 10^{-4}$ | $7.712 \times 10^{-5}$ | $1.714 \times 10^{-6}$ |
| Fe 56 | $4.118 \times 10^{-3}$ | $3.019 \times 10^{-3}$ | $1.503 \times 10^{-3}$ | $2.759 \times 10^{-5}$ |
| Rb 87 | $2.372 \times 10^{-8}$ | $2.166 \times 10^{-8}$ | $9.925 \times 10^{-9}$ | $5.467 \times 10^{-10}$ |
| Sr 88 | $5.785 \times 10^{-7}$ | $4.263 \times 10^{-7}$ | $1.402 \times 10^{-7}$ | $5.333 \times 10^{-9}$ |
| Y 89  | $1.479 \times 10^{-7}$ | $1.150 \times 10^{-7}$ | $3.771 \times 10^{-8}$ | $1.540 \times 10^{-9}$ |
| Zr 90 | $1.622 \times 10^{-7}$ | $1.272 \times 10^{-7}$ | $4.202 \times 10^{-8}$ | $1.652 \times 10^{-9}$ |
| Zr 96 | $3.767 \times 10^{-9}$ | $3.968 \times 10^{-9}$ | $1.871 \times 10^{-9}$ | $8.409 \times 10^{-11}$ |
| Ba 138| $1.505 \times 10^{-7}$ | $2.570 \times 10^{-7}$ | $1.310 \times 10^{-7}$ | $6.036 \times 10^{-9}$ |
| Pb 208| $3.108 \times 10^{-8}$ | $3.635 \times 10^{-8}$ | $1.043 \times 10^{-7}$ | $9.500 \times 10^{-8}$ |

**Listing 1.** Code example to read NuGrid's 2M$_\odot$, Z = 0.02 with a python notebook on the WENDI Astrohub platform model and make an abundance profile plot.

```
\vspace{-12pt}
%pylab ipympl
from nugridpy import nugridse as~mp

data_dir1 = ''/data/ASDR/NuGrid''
mp.set_nugrid_path(data_dir1)
pt = mp.se(mass=2,Z=0.02)

use_updated_model = True
if use_updated_model:
    data = ''/data/nugrid/data/set1upd''
    data_dir2 = data +''/set1.1/ppd_wind/RUN_set1upd_m2z1m2/H5_out''
    pt = mp.se(data_dir2)

species = ['H-1','He-4','C-12','C-13','N-14','O-16']
ifig = 108; close(ifig); figure(ifig)
pt.abu_profile(isos=species, ifig=ifig, fname=18000,
logy=True)
ylim(-7,0); xlim(0.551,0.555)
```

## 5. Conclusions and Future Plans

In recent years, the NuGrid collaboration computed stellar models describing the full evolution and nucleosynthesis of stars covering a wide range of initial masses and metallicities. These models were presented in different publications. For AGB stars, we refer to [47–49,51]. In this work, we defined the NuGrid AGB evolution and nucleosynthesis data set, specifying for every specific combination of mass and metallicity the reference providing the most up-to-date stellar model. Moreover, in this work we updated the *s*-process nucleosynthesis of low-mass AGB models with initial mass M = 2 and 3M$_\odot$ at around-solar metallicity, adopting the recent $^{22}$Ne + $\alpha$ rates from [59]. This resulted in an improvement of the comparison between our theoretical prediction and key observables, such as isotopic ratios in presolar SiC grains and surface abundances of Ba stars.

This preliminary AGB data set still needs some major extensions in order to provide all the essential inputs for GCE studies. In particular, low-mass AGB models and related nucleosynthetic ejected yields need to be computed for the lowest metallicity of the grid (Z = 0.0001), the models with initial mass M = 1 and 1.65M$_\odot$ and for all metallicities will also need to be recalculated using the more recent $^{13}$C-pocket setup by [33], while the masses M = 4, 5, 6, 7M$_\odot$ need to be calculated for super-solar metallicity at Z = 0.03 and recomputed for all metallicities adopting the recent $^{22}$Ne + $\alpha$ rates from [59].

Finally, the $^{12}$C/$^{13}$C ratios in our Z = 0.0001 models is in the order of 1000, while CEMP-*s* stars show $^{12}$C/$^{13}$C ratios less than solar, often in the range 5 to 30. This is a clear indication to missing physics either in the lowest Z AGB models, the assumed CEMP-s star scenario or both. A dedicated effort to solve this issue, for example by including a physics-based prescription of extra-mixing in AGB stars, is one of the main goals for the future.

**Author Contributions:** Conceptualization, U.B. and M.P.; methodology, U.B. and M.P.; software, U.B. and A.T.; validation, U.B.; formal analysis, U.B.; investigation, U.B., M.P., A.T., P.D. and F.H.; resources, U.B.; data curation, U.B. and F.H.; writing—original draft preparation, U.B. and M.P.; writing—review and editing, P.D., F.H. and A.T.; visualization, U.B.; supervision, U.B. and M.P; project administration, U.B. and M.P.; funding acquisition, U.B. and M.P. All authors have read and agreed to the published version of the manuscript.

**Funding:** This work is supported by the European Union's Horizon 2020 research and innovation programme (ChETEC-INFRA—Project no. 101008324), and by STFC (through the University of Hull's Consolidated Grant ST/R000840/1). U.B. and M.P. thank the University of Hull High Performance Computing Facility for providing ongoing access to viper. We acknowledge the financial support of NuGrid/JINA-CEE (NSF Grant PHY-1430152), and the support from the IReNA network (US NSF AccelNet) and the ChETEC COST Action (CA16117, European Cooperation in Science and Technology). M.P. thanks the ERC Consolidator Grant (Hungary) programme (RADIOSTAR, G.A. n. 724560) and the "Lendület-2014" Programme of the Hungarian Academy of Sciences (Hungary) for their support. F.H. and P.D. give thanks for the support of the NSERC, under award SAPPJ-2021-00032 for the project Nuclear Physics of the Dynamic Origin of the Elements, carried out by the Canadian Nuclear Physics for Astrophysics Network (CANPAN).

**Informed Consent Statement:** Not applicable.

**Institutional Review Board Statement:** Not applicable.

**Data Availability Statement:** The data presented in this study are openly available in NuGrid repository, as specified at https://nugrid.github.io/content/data (accessed on 1 March 2022).

**Acknowledgments:** This research was enabled in part by support provided by WestGrid (www.westgrid.ca, accessed on 1 March 2022) and Compute Canada Calcul Canada (www.computecanada.ca, accessed on 1 March 2022). The computing facilities (in particular the Orcinus cluster) were mostly supported by the UBC Chemistry Department Team led by Mark Thachuk. NuGrid data is served by Globus (www.globus.org, accessed on 1 March 2022).

**Conflicts of Interest:** The authors declare no conflict of interest.